\documentclass[9pt,conference]{IEEEtran}
\usepackage{graphicx} 
\usepackage{url}
\usepackage{amsmath}
\usepackage{xspace}
\usepackage{hyperref}
\title{SsgCaps: A controlled dataset for the evaluation of sound scene generation algorithms}
\author{%
     Modan Tailleur$^1$, 
     Junwon Lee$^2$,  Laurie M. Heller$^3$,Mathieu Lagrange$^1$, \\
     Keunwoo Choi$^4$, Brian McFee$^5$, Keisuke Imoto$^6$, Yuki Okamoto$^7$
}

\newcommand\open{\emph{SsgCaps}\xspace}
\newcommand\closed{\emph{SsgCapsClosed}\xspace}

\renewcommand{\baselinestretch}{1}

\begin{document}
\maketitle

\begin{abstract}

Sound Scene Generation is about the automatic synthesis of artificial sound scenes. We introduce SsgCaps, a publicly available dataset of human-engineered sound scenes wherein each scene matches a precisely structured prompt that guides the sampling process. The corresponding prompts are sampled from a predefined action-based typology that allows extensive sampling while retaining plausibility. SsgCaps is a sound scene dataset derived from the unpublished reference dataset for Task 7 of the 2024 DCASE Challenge edition, which contained private- and public-domain audio samples. In contrast, SsgCaps contains only public-domain audio samples, allowing us to open this dataset to the community. 

To make this dataset useful to the community, we first elaborate on the rationale for the prompt and dataset structure. We then perform a comparative quantitative analysis of the 2 versions of the dataset. To do so, we compare both versions to the audio synthesized by the SSG algorithms submitted to the challenge using Fréchet Audio Distance (FAD) and Kernel Audio Distance (KAD) as well as perceptual ratings. This analysis shows only small differences, which enables us to recommend the open version for further benchmarking of SSG algorithms. 

\end{abstract}

\begin{IEEEkeywords}
sound scene generation, open dataset, audio processing, machine learning
\end{IEEEkeywords}

\section{Introduction}

Sound scene generation (SSG) is helpful for various kinds of tasks ranging from classification and detection to sonification.
Early applications of SSG aimed to help design better algorithms and experimental protocols rather than to produce sounds for human listeners. To our knowledge, SSG for machine listening assessment was introduced in the inaugural edition of DCASE for the design of a controlled data set to evaluate the detection of audio events \cite{stowell2015detection}. Compared to using annotated recordings, the benefits of using a controlled dataset generated using SSG are twofold. First, a large amount of data can be generated, and second, there is less uncertainty in the ground truth labels \cite{lafay2016morphological, salamon2017scaper}. Furthermore, provided that the data set has been stratified in a meaningful way, interesting outcomes of the behaviors of the evaluated systems can be gathered \cite{lafay2017sound}.

With the rise of deep learning and its ever-increasing need for training data, SSG is also used for Training Set Synthesis (TSS)  \cite{gontier2021polyphonic}. Simply sequencing different recorded samples in various ways proved to be highly effective in helping stabilize training and inducing invariance to overlapping sound events and other confounding factors that can be easily simulated, such as propagation and room acoustics. Although TSS can potentially be used for initial training, it is probably more effective as a way to fine-tune existing models for specific tasks with low resource requirements. An example would be to customize a sound source detector for a given city soundscape by using a few recordings of the specific sound sources of the city where the sensor network will be deployed.

 With respect to TSS, one limitation of early SSG tools such as SimScene\footnote{\url{https://bitbucket.org/mlagrange/simscene/src/master}} or scaper\footnote{\url{https://github.com/justinsalamon/scaper}} is that they require the availability of a diverse set of audio recordings of individual events. In order for SSG to be appropriate for end-user applications such as audio track generation for movies and computer games, it would be more convenient to prompt the system with a text-based "prompt", i.e., a description of the sound scene.

To our knowledge, the first use of deep learning for building an SSG is presented in \cite{kong2019acoustic}, a system that is prompted using only one word, providing a very loose description of the acoustic content. Trained using the DCASE 2016 Task 1 acoustic scene dataset, the system required a word to be chosen from the following typology: \emph{Bus, Cafe, Car, City center, Forest path, Grocery store, Home, Lakeside beach, Library, Metro station, Office, Residential area, Tram and Urban park}. 

Starting from this point, the literature evolved to use less constrained prompts, fueled by the availability of audio captioning datasets, such as CLOTHO \cite{drossos2020clotho}, AudioCaps \cite{kim2019audiocaps}, and WavCaps \cite{mei2024wavcaps}. Interestingly, sample sequencing SSG systems such as SimScene and scaper can be seen as tackling the inverse task of sound event detection (SED), as the output of the SED system that is a transcript of sound event occurrences is all that is needed to feed sample sequencing SSG systems. Conversely, prompt-based SSGs are, in some sense, tackling the inverse task of audio captioning. 

While recorded audio that was captioned with free-form descriptions makes it possible to gather a large amount of data for training deep-learning-based SSGs, we believe that for evaluation purposes, it is preferable to rely on controlled datasets in which we sample the prompts from a fixed typology first and then handcraft corresponding audio scenes. Indeed, the above-cited audio caption datasets have no constraints on the syntax of the caption, which can lead to some consistency issues when conducting systematic performance analysis. Prompts found in AudioCaps such as "She is talking" or "Dog continuously barking while middle aged woman vocalizes in distinct accent and laughs" are problematic as they are very ambiguous.

To provide solid ground for evaluating the merit of SSG systems, task 7 of the 2024 DCASE Challenge edition \cite{lee:hal-04794208} focused on generating realistic environmental audio scenes from controlled text prompts. Participants had to create models that took a text description specifying both a foreground sound and a background context (e.g., “a dog is barking with water in the background”) and produced a 4-second audio waveform corresponding to that description. A key characteristic of the evaluation set of this task is that the prompts were manually crafted by the organizing team to be semantically unambiguous and balanced across categories. The corresponding audio recordings were then produced by a professional sound designer to ensure the highest possible audio quality and accuracy of prompt-sound matches. The first version of the dataset included some private audio samples. This first version will be termed \closed in the remainder of the paper. 

The goal of this paper is to enable the community to effectively utilize a balanced high-quality SSG dataset, which requires more than simply making the sounds available on a repository. This requirement motivates the three main contributions of this paper: 1) to introduce a second version of this dataset, \open, that replaces sounds originally engineered using private audio libraries with sounds engineered solely from Freesound recordings, allowing us to make \open publicly available\footnote{\url{https://zenodo.org/records/15630417}}, 2) to demonstrate that this new version can be used equivalently for further evaluation of SSG systems, and 3) to introduce the rationale and unique features of the dataset's design.

The remainder of the paper is organized as follows: Section \ref{sec:prompt} details the process of generating the prompts of \open. Section \ref{sec:dataset} describes some statistics that characterize the content of \open. Section \ref{sec:Qualitative} describes and compares the perceptual ratings of the sounds in \open and \closed. In Section \ref{sec:analysis}, we perform a quantitative analysis comparing \open to \closed with both Fréchet Audio Distance (FAD) \cite{kilgour2018fr} and Kernel Audio Distance (KAD) \cite{chung2025kad}. 

\begin{figure*}[th]
    \centering
    \includegraphics[width=.8\linewidth]{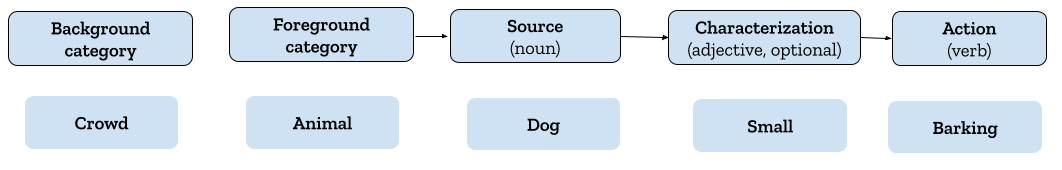}
    \caption{The prompts are generated using a fixed typology composed of the several elements presented on top. Each item is instantiated (bottom) to build the following prompt: "A small dog is barking in a crowd".}
    \label{fig:typology}
\end{figure*}

\section{Design of prompts}
\label{sec:prompt}

The prompts in \open are identical to those in \closed, and their design was thus originally introduced in the DCASE Challenge 2024. Herein we review the design of the dataset while introducing  new information about the full process and rationale behind the design and highlighting its most desirable features. 

\subsection{Prompt structure}

Each prompt follows the following structure:
”\textit{Foreground} with \textit{Background} in the background,”
to specify both the primary sound source and its acoustic context separately. As shown on Figure \ref{fig:typology}, the primary sound is then described by:
\begin{enumerate}
    \item its foreground category;
    \item the specific source within this category;
    \item a characterization of this source to describe it more precisely if needed;
    \item the action of the source that created some audible sound.
\end{enumerate}


The dataset is thus organized  into two main categories:
foreground and background sounds. The foreground category encompasses six distinct types of sounds: \emph{Animal, Vehicle, Human,
Alarm, Tool, and Entrance sounds}. These were chosen to represent a diverse range of common sound sources in everyday environments. For background sounds, the dataset includes five categories:
\emph{Crowd, Traffic, Water, Birds, and Presence} sounds. Presence is the recorded background "silence" of a quiet location when no dialogue is spoken and no sound source can be perceived \cite{holman2013sound}.

This categorization enables the creation of realistic sound scenes with clear foreground-background relationships.

\subsection{Generation of Prompts} 

\begin{table}
    \caption{Prompts for one source of each main foreground category.}
    \label{tab:prompts}
    \centering
    \begin{tabular}{l|l}
    \hline
    Main category & Prompts\\
    \hline
       human &	a child is coughing\\
&	a child is screaming\\
&	a child is laughing\\
&	a child is crying\\
 \hline
animal&	a cat is meowing\\
&	a cat is purring\\
& 	a cat is hissing\\
&	a cat is wailing\\
 \hline
vehicle&	a race car is idling\\
&	a race car is passing by\\
&	a race car is starting\\
&	a race car is speeding up\\
 \hline
alarm&	a doorbell is ringing\\
&	a bicycle bell is ringing\\
&	a school bell is ringing\\
 \hline
tool&	a small gun is reloading\\
&	a small gun is shooting\\
 \hline
entrance&	a car door is opening\\
&	a car door is closing\\
&	a car door is being slammed\\
\hline
    \end{tabular}
\end{table}

Before collecting sounds, a list of 360 audio/caption couples with prototypical prompts was generated and refined for foreground sound sources. The sound sources and their actions had to be unambiguously described by an adjective + noun and verb (e.g., "a small dog is barking"). In this example, a bark from a small dog sounds higher-pitched than a bark from a big dog. However, adjectives and/or actions that did not have obvious acoustic consequences (e.g. "a red bird is perching") were excluded. The most specific noun that was conveyed by the sound was used, e.g. "dog" was used instead of "animal," but more general nouns were used if the sound did not reveal details about the source (e.g. for clapping sounds, the prompt was "a human is clapping"). Some examples of prompts for each main foreground category are given in Table \ref{tab:prompts}.  Sounds with speech and music were not included, but non-verbal vocalizations were allowed (e.g. "a person is snoring"). An effort was made to pair different types of foreground objects with the same actions, such as a race car and a motorcycle speeding up, but only if the foreground sources were deemed to sound distinct to a general audience. For example, the general term "motor vehicle" was used with the verb "crashing" because it was not obvious that the crash sound would specify much about the vehicle itself, whereas both "a small car" and "a bus" were used with "is passing by" because their sounds are distinct. After the list within each category was refined to contain multiple examples of each foreground sound source paired with different actions, sound samples were obtained and edited by our team's sound designer. The organizers then listened to the sound samples and rated them based on how easily identifiable and distinct the sources and actions were, how clear the audio quality was, and whether there were competing sounds. When a problem was discovered, a new sound sample was found. Revisions were iterated until a final set of 310 audios was selected.
Our background sounds of traffic, birds, crowd noise, and flowing water were chosen to have reasonably continuous textures that would be distinguishable from the foreground sounds without being distracting. 

\section{Dataset description} \label{sec:dataset}

\subsection{Dataset split}

The dataset contains 310 prompts with corresponding audio files. Between 10 and 12 exemplars per type of background are present for each foreground category, as shown in Table \ref{tab:foreground}. There are 60 audio files in the development set and 250 in the evaluation set, distributed approximately equally among the foreground categories, as shown in Table \ref{tab:foreground_dev_eval}. As shown in Table \ref{tab:backgroun_dev_eval}, two background types were withheld from the development set and only present in the evaluation set.

\subsection{Audio}

Maximizing audio quality, we carefully selected audio recordings freely available from the Freesound database \cite{akkermans2011freesound}. Those recordings are then edited for creating the mixtures corresponding to the prompts. Key constraints for the resulting mixtures include:
\begin{itemize}
    \item 4-second 16-bit mono audio snippets at 32 kHz sampling rate;
    \item No music;
    \item No intelligible speech.
\end{itemize}
The background and foreground were mixed without any post-processing effects such as compression or reverb. The level of the background was adapted per frame of about one second to maintain the salience of the foreground throughout the entire mixture. The original recordings are each exclusively used for one prompt and thus are not shared between development and evaluation sets.


\begin{table}[h]
\caption{Number of each background type in each category, for the whole dataset.}
\label{tab:foreground}
\centering
\begin{tabular}{lrrrrr}
\hline
\textbf{Category} & \textbf{No back.} & \textbf{Birds} & \textbf{Crowd} & \textbf{Traffic} & \textbf{Water} \\
\hline
alarm    & 10 & 10 & 10 & 10 & 10 \\
animal   & 10 & 11 & 10 & 10 & 11 \\
entrance & 10 & 10 & 10 & 10 & 10 \\
human    & 11 & 12 & 11 & 11 & 11 \\
tool     & 10 & 10 & 10 & 10 & 10 \\
vehicle  & 11 & 10 & 11 & 10 & 10 \\
\hline
\end{tabular}
\end{table}

\begin{table}[h]
\caption{Number of elements per category in development, evaluation and perceptual evaluation sets.}
\label{tab:foreground_dev_eval}
\centering
\begin{tabular}{lrrrr}
\hline
\textbf{Category} & \textbf{Dev. set} & \textbf{Eval. set} &\textbf{Perceptual Eval. set} & \textbf{Total} \\
\hline
alarm    & 10 & 40 & 4& 50 \\
animal   & 10 & 42 & 4& 52 \\
entrance & 10 & 40 & 3& 50 \\
human    & 10 & 46 & 3& 56 \\
tool     & 10 & 40 & 4& 50 \\
vehicle  & 10 & 42 & 3& 52 \\
\hline
\end{tabular}
\end{table}

\begin{table}[h]
\caption{Number of elements per background type in development and evaluation sets.}
\label{tab:backgroun_dev_eval}
\centering
\begin{tabular}{lrrr}
\hline
\textbf{Background} & \textbf{Dev. set} & \textbf{Eval. set} & \textbf{Total} \\
\hline
No background       & 0  & 62 & 62 \\
birds    & 0  & 63 & 63 \\
crowd    & 19 & 43 & 62 \\
traffic  & 22 & 39 & 61 \\
water    & 19 & 43 & 62 \\
\hline
\end{tabular}
\end{table}

\subsection{Difference between \closed and \open}

The main issue with \closed is that some audios are composed of sounds originally engineered using private audio libraries, preventing us from publicly disclosing it. In \open, those audios are replaced with sounds engineered from Freesound recordings, allowing us to make \open publicly available.  Overall, 24 of the 360 samples in the \closed\ dataset are privately owned and therefore are herein replaced with sounds designed exclusively from Freesound sources. The prompts are unchanged.

\section{SsgCaps evaluation}

In this section, we evaluate the impact of using \open as a reference dataset for the computation of objective metrics in the evaluation of SSG systems. As a preliminary step, we verify in Section~\ref{sec:Qualitative}  that the perceptual ratings from the DCASE 2024 Challenge remain usable after removing the ratings of the sounds that are no longer present in \open. Based on this verification, we leverage those ratings in Section~\ref{sec:analysis} to assess the impact of using \open instead of \closed as a reference for objective evaluation metrics. 

\subsection{Qualitative analysis} \label{sec:Qualitative}

Full details of the perceptual rating method and participants for the evaluation of \closed in the original DCASE 2024 Challenge are summarized herein and detailed in \cite{lee:hal-04794208}. Perceptual ratings were made on an absolute scale from 0 to 10 in which 10 was the best possible sound imaginable. Each sound scene was played while its prompt was displayed on the computer screen. Listeners rated the sound twice within each trial to indicate how well it matched the foreground and background components of the prompt. This allowed us to control the relative weight of foreground and background in the Final Rating score, which could not be done if each prompt received only one overall rating. After 144 sounds from 6 systems (24 per system) were rated in random order, all sounds were repeated in random order to obtain a perceptual quality rating. No prompt was given during quality ratings so that the raters would focus on sound clarity/distortion rather than prompt fit.  We made an \textit{a priori }decision to form a final rating from a 2:1:1 weighting of foreground: background: quality, emphasizing the foreground fit for our sound scene generation task. 

\begin{figure}[t]
    \centering
    \includegraphics[width=.8\linewidth]{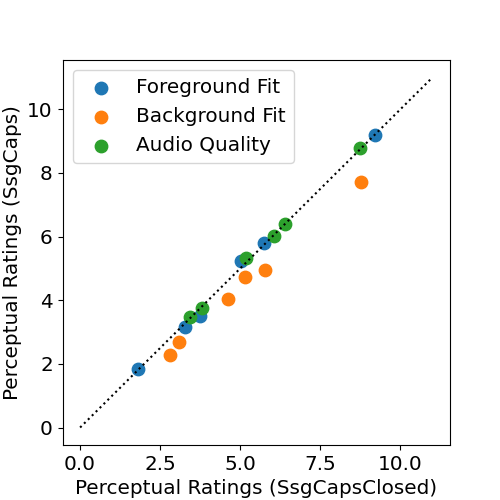}
    \caption{Perceptual ratings of sounds from each of the 6 systems that were perceptually evaluated (eval, baseline, and four state-of-the-art systems) compared between the \closed and the \open datasets. Perceptual ratings were for: foreground fit, background fit, and audio quality.}
    \label{fig:perceptual}
\end{figure}

As explained in Section \ref{sec:dataset}, in \open, 24 privately-owned sounds were removed from \closed and replaced by sounds designed from Freesound sources. In order to check if the conclusions reached by taking into account the perceptual ratings are still valid for \open, we conducted the following experiment. We performed a virtual perceptual evaluation of \open by considering the ratings of the 141 remaining sounds, leaving the perceptual ratings of 3 sounds, one from each of the categories of human, entrance, and vehicle sounds. Comparing the results of the perceptual experiment of \closed and the ones of the virtual experiment of \open, the rankings of each system remained the same in SSG-Open as in the original DCASE competition. As shown in Figure \ref{fig:perceptual}, the perceptual evaluation of the two datasets is quite similar, with a slight improvement for \open indicating that the removed sounds were probably harder to imitate, perhaps due to their higher recording quality.

\subsection{Quantitative analysis} \label{sec:analysis}

Quantitative evaluation in SSG is routinely performed using the Fréchet Audio Distance (FAD) \cite{kilgour2018fr}. The Fr\'echet distance between the two distributions $r$ and $t$ of sounds projected in a given embedding space is calculated as follows: 

\begin{equation}\label{eq:FAD}
\text{FAD}(r, t) = \left\| \mu_r - \mu_t \right\|_2 + \text{tr} \left( \Sigma_r + \Sigma_t - 2\sqrt{\Sigma_r \Sigma_t} \right)
\end{equation}
where $\mu_x$ and $\Sigma_x$ are, respectively, the mean and covariance matrix of a given distribution $x$. The FAD calculation compares the two datasets in terms of fit to domain with the comparison of means, but also in terms of diversity by including a form of covariance comparison in the equation. Initially performed using the VGGish embedding, the FAD computed using the PANNs embeddings \cite{tailleur2024correlation} allows for a better alignment of the metric with human perceptual ratings. Removing the Gaussian assumption of the FAD can be achieved by considering the Kernel Audio Distance (KAD)  \cite{chung2025kad}:

\begin{equation}
\begin{aligned} \text{KAD}= & \frac{1}{n(n-1)} \sum_{i \neq j} k\left(x_{i}, x_{j}\right)+\frac{1}{m(m-1)} \sum_{i \neq j} k\left(y_{i}, y_{j}\right) \\ & -\frac{2}{n m} \sum_{i=1}^{n} \sum_{j=1}^{m} k\left(x_{i}, y_{j}\right)\end{aligned}
\end{equation}
where $X=\left\{x_{i}\right\}_{i=1}^{n}$ is the finite samples in the reference set and $Y=\left\{y_{i}\right\}_{i=1}^{n}$ is the finite samples in the evaluated set.
The kernel $k$ is chosen as a radial basis function:
\begin{equation*}
k(\mathbf{x}, \mathbf{y})=\exp \left(-\frac{\|\mathbf{x}-\mathbf{y}\|^{2}}{2 \sigma^{2}}\right)    
\end{equation*}
where $\sigma$ is the median  pairwise distance between the embeddings within the reference set.

\vspace{0.5em}
\noindent\textbf{Comparison with Perceptual Ratings}

\begin{figure}[t]
    \centering
    \includegraphics[width=.9\linewidth]{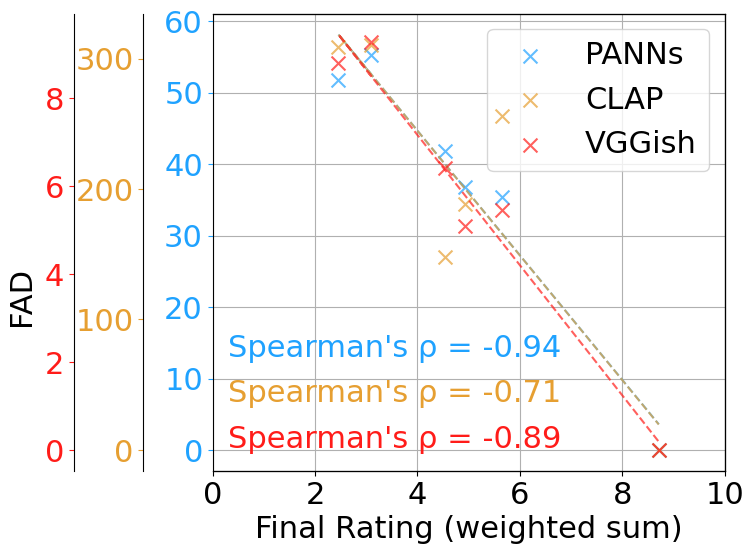}\\\includegraphics[width=.9\linewidth]{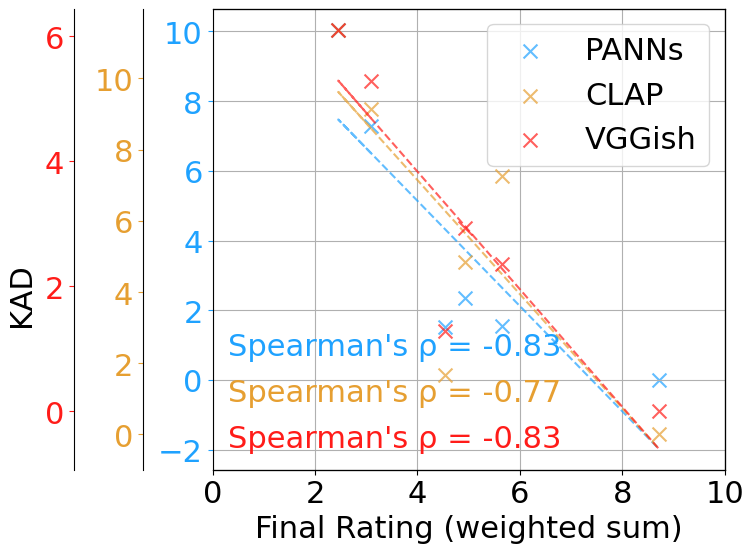}
    \caption{FAD (top) and KAD (bottom) computed among the audio generated by the submitted systems and the \open dataset versus the perceptual evaluation data relevant for the \open dataset.}
    \label{fig:pr}
\end{figure}

In Figure \ref{fig:pr} we display the FAD and KAD computed from the audio generated by the submitted systems and the \open dataset versus the perceptual evaluation data relevant for the \open dataset. It shows a rather high level of correlation that is extremely similar to the one achieved with the \closed dataset \cite{lee:hal-04794208}.

\vspace{0.5em}
\noindent\textbf{Comparison with \closed}

In \cite{lee:hal-04794208}, an extensive benchmarking of the performance of 18 SSG systems is conducted by comparing their respective performance using PANNs-based FAD with \closed as the reference dataset.

\begin{figure}[t]
    \centering
    \includegraphics[width=.5\linewidth]{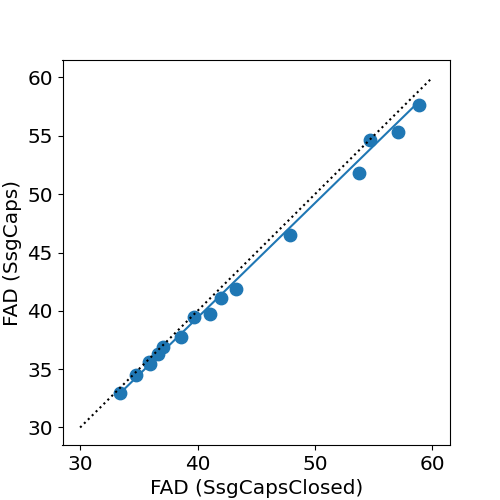}\includegraphics[width=.5\linewidth]{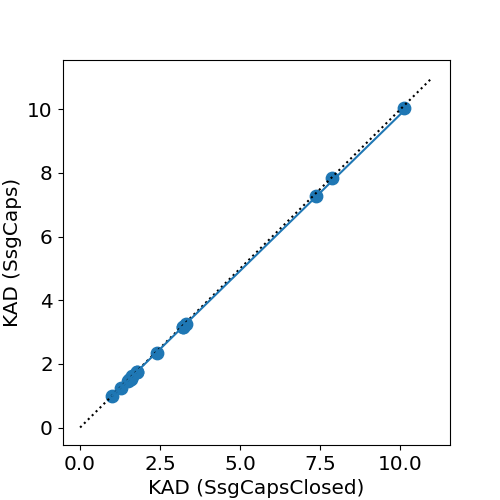}
    \caption{FAD (left) and KAD (right) computed among the audio generated by the 18 state-of-the-art systems and the \open dataset versus the \closed dataset.}
    \label{fig:closed}
\end{figure}

By computing the FAD and KAD on both \open and \closed  datasets with respect to the audio generated by the 18 systems, we can gauge the level of similarity among the two versions of the dataset. Figure \ref{fig:closed} shows two scatter plots, respectively, illustrating the FAD and the KAD  where we can observe that each linear regression is quite close to the diagonal, indicating high similarity.


\section{Conclusion}

This paper introduced the \open dataset. It provides some resources to evaluate the performance of SSG systems. We believe that the structure of the prompts allows for a good balance between expressivity and coherence while the emphasis on high sound quality enhances its validity. We hope that providing the rationale and details of our method for generating the prompts and sounds will be useful to other dataset developers with similar goals. Compared to the \closed dataset, the qualitative and quantitative evaluation does not show any important changes, allowing us to recommend the \open dataset for further benchmarking of SSG systems.



\end{document}